\documentclass[twocolumn]{aastex631}
\usepackage{times}
\usepackage{amsmath}

\shorttitle{HarD X-rays in Cen~A}
\shortauthors{Rani et al.}

\begin{document}
\title{Hard X-ray emission in Centaurus~A}

\author[0000-0001-5711-084X]{B. Rani}
\affiliation{Korea Astronomy and Space science Institute, 776 Daedeokdae-ro, Yuseong-gu, Daejeon 30455, Korea}
\affiliation{University of Science and Technology, Korea, 217 Gajeong-ro, Yuseong-gu, Daejeon 34113, Korea}
\affiliation{Department of Physics, American University, Washington, DC 20016, USA}

\author{S.~A.~Mundo}
\affiliation{Department of Astronomy, University of Maryland, College Park, MD 20742, USA} 

\author{R.~Mushotzky}
\affiliation{Department of Astronomy, University of Maryland, College Park, MD 20742, USA} 

\author{A.~Y.~Lien}
\affiliation{University of Tampa, Department of Chemistry, Biochemistry, and Physics, 401 W. Kennedy Blvd, Tampa, FL 33606, USA}

\author{M.~A.~Gurwell}
\affiliation{Center for Astrophysics | Harvard \& Smithsonian, 60 Garden Street, Cambridge, MA 02138 USA}

\author[0000-0001-8229-7183]{J.~Y.~Kim}
\affiliation{Department of Astronomy and Atmospheric Sciences, Kyungpook National University, Daegu 702-701, Republic Of Korea}
\affiliation{Max-Planck-Institut f{\"u}r Radioastronomie, Auf dem H{\"u}gel 69, D-53121 Bonn, Germany}

%*****************************************************************************
\begin{abstract}
We used 13~years of {\it Swift}/BAT observations to probe the nature and origin of hard X-ray (14-195~KeV) emission in Centaurus~A.  Since the beginning of the Swift operation in 2004, significant X-ray variability in the 14-195~KeV band is detected, with mild changes in the source spectrum. Spectral variations became more eminent after 2013, following a softer-when brighter trend. Using the power spectral density method, we found that the observed hard X-ray photon flux variations are consistent with a red-noise process of slope, $-1.3$  with no evidence for a break in the PSD. We found a significant correlation between hard X-ray and 230~GHz radio flux variations, with no  time delay longer than 30~days. The temporal and spectral analysis rules out the ADAF (advection-dominated accretion flow) model, and confirms that the hard X-ray emission is produced in the inner regions of the radio jet. 
\end{abstract}
%
%%%%%%%
%
\keywords{galaxies: active; galaxies: individual (Centarus~A); X-rays: galaxies; 
radio continuum: galaxies}
%*****************************************************************************

%*****************************************************************************
\section{Introduction}
The X-ray emitting sites in Active Galactic Nuclei (AGN)  are not
well understood. 
X-rays could either originate in the immediate vicinity of the central black hole 
(disk/corona) or further out in the jets. Some of these X-rays penetrate into the disk, 
where they are re-processed to produce {\it the 'reflection' spectrum} that includes 
the Fe~K$\alpha$ line \citep{lohfink2013, hinkle2021}.
The geometry of the disk/corona  is an active area of research. 
A detailed understanding of disk/corona/jet contribution in the observed X-ray emission 
is a critical element in unraveling  how the central engine of an AGN operates and feeds 
the jet. In this paper, we investigated the origin of hard X-ray emission in a nearby AGN, 
Centaurus A (Cen~A hereafter), using the observed variations in the X-ray and radio regimes.

At a distance of $d\simeq3.8$\,Mpc \citep{harris10}, Cen~A is the closest AGN hosting a 
supermassive black of $\sim5\times10^{7}M_{\odot}$ \citep{neumayer2010}.  
From the radio morphology of the lobes, Cen~A is classified as Fanaroff-Riley type I 
\citep{fanaroff74}. In fact, Cen~A jet has been detected and extensively studied across the 
whole electromagnetic spectrum, 
radio to $\gamma$-rays \citep{wykes2015, muller2014, hardcastle2003, worral2008, abdo2010, eht2021}. 
In 2004, the source was first detected at TeV energies by H.E.S.S. \citep[High Energy Stereoscopic System,][]{hess2009},  and later by the Fermi/LAT at GeV energies \citep{abdo2010}. 
Spatial extension of $\gamma$-ray emission is detected both at GeV \citep{abdo2010} and 
TeV energies \citep{hess2020}, the physical origin of which remained unclear.

\begin{figure*}
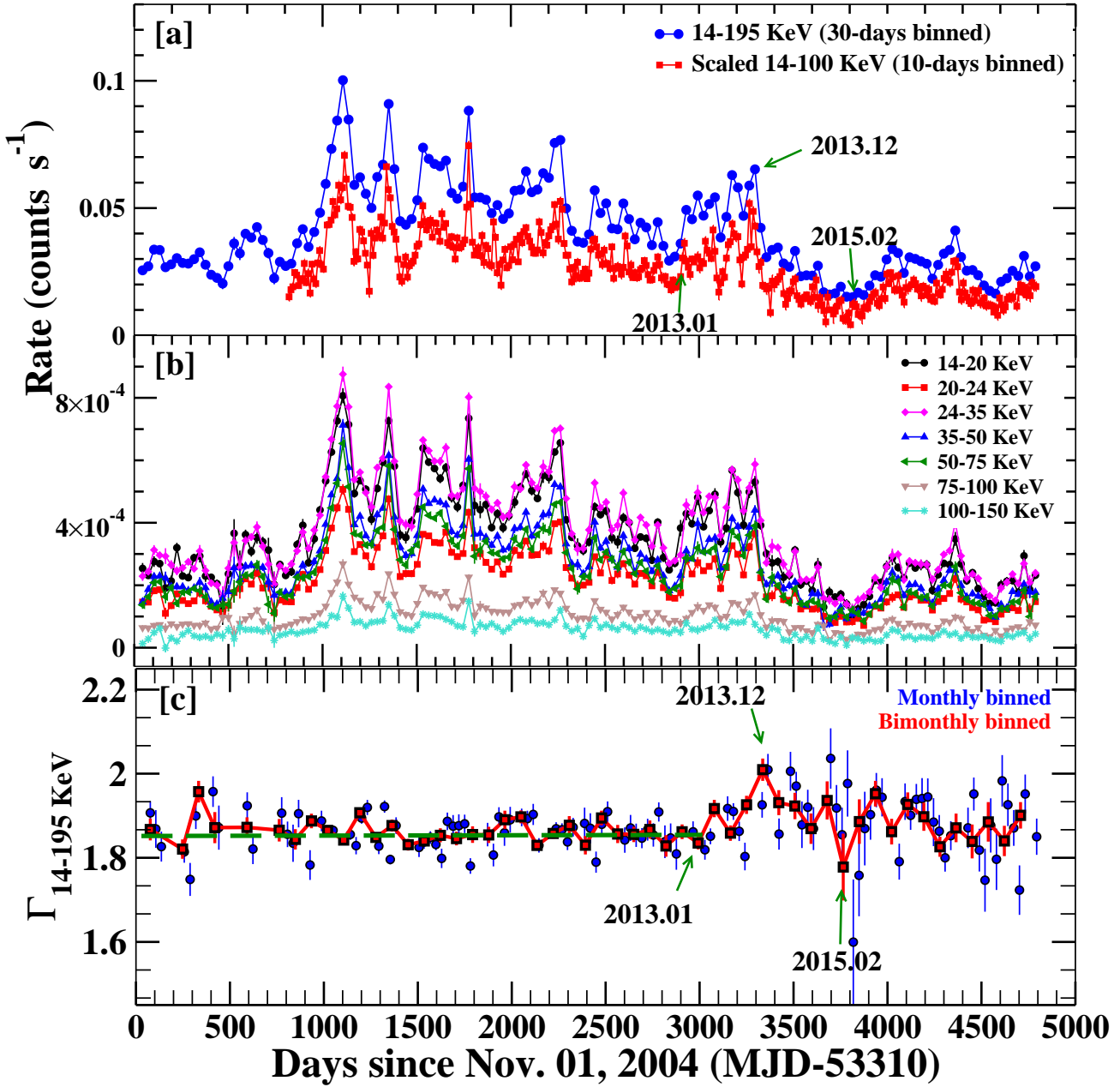

\includegraphics[scale=0.68, trim=0 48.9 0 0, clip]{fig1_new.eps}
\includegraphics[scale=0.71, trim=-14 0 0 2.2, clip]{XIndx.eps}
\caption{Photon flux and spectral variations observed in Cen~A since November 2004: 
[a] Monthly averaged 14-195~KeV  (blue circles) and 10-day binned 14-100~KeV (red squares) light curves. The 10-day binned 
data is scaled by a factor of 15, [b]  Monthly averaged light curves in different energy bins, and [c] 
Hard X-ray photon index variations observed in the source. The green arrows mark the prominent spectral variability 
phases of the source (see Section 2.1 for more details). }
\label{fig1}
\end{figure*}

The X-ray (0.1-7~KeV) morphology of Cen~A consists of a  central bright AGN and  a faint  
jet component surrounded by diffuse emission \citep{kraft2002}. 
The source has a complex X-ray spectrum, comprising  a soft (0.1-2~KeV) thermal plasma, a  power-law continuum, and strong absorption of the power-law continuum. The location and structure of the absorbing material is still under 
debate \citep[e.g.][]{evans2004, markowitz2007, fukawaza2011}. 
The hard X-ray spectrum of the source can be well described by an 
absorbed power-law or thermal Comptonization spectrum with an Fe~K$\alpha$ line, with 
no evidence for a high-energy  exponential rollover \citep{furst2016}. Detection of a 
weak reflection component has been reported \citep{fukawaza2011, burke2014}, 
however, recent analysis has placed a very tight upper limit on the  presence of a reflection component \citep{beckmann2011, furst2016, rothschild2011}. 
Small changes in the  hard X-ray power-law continuum photon index have been reported over the past decades \citep{baity1981, rothschild2011, furst2016}, with the slope being bounded between 1.6-1.85. This range of indices is consistent with what is found for  Seyfert
galaxies.

While the continuum flux is strongly variable over time, the flux of 
iron line remained stable, indicating a strong variability of the equivalent width of the 
iron line \citep{rothschild2006}.
Even the joint spectral analysis using truly simultaneous {\it XMM-Newton} and {\it NuSTAR} data could 
not determine the physical origin of the hard X-ray emission in the source \citep{furst2016}. 
The study found no significant contribution from the hot interstellar medium (ISM), the outer jet and off-nuclear 
point sources, in the hard X-ray spectrum. Lack of reflection rules out the standard Seyfert-like geometry of the source of hard X-rays and reprocessing material. Comptonization in an advection dominated accretion flow (ADAF)   
or at the base of the inner jet or a combination of the two  were proposed as the possible mechanisms 
 for the hard X-ray emission in the source \citep{furst2016}.

We present here a comprehensive analysis of the observed 
variations in  X-ray and radio regions to better understand the nature and origin of 
the hard X-ray emission. The paper is structured as follows. In section 2, we present 
the data analysis and results. Results are discussed in Section 3, and summarized in Section 4.

\section{Data  analysis and Results}

\subsection{X-rays}
We investigated the X-ray flux and spectral variations of the source using data from the 
{\it Neil Gehrels Swift Observatory}/Burst Alert Telescope ({\it Swift}/BAT)  157-Month Hard X-ray Survey\footnote{https://swift.gsfc.nasa.gov/results/bs157mon/671}.  
While in survey mode (not specifically targeting a GRB), BAT continuously scans the sky with a time resolution 
as fine as 64~s \citep{krimm2013}.  Monthly averaged light curves and spectra of sources in 
the hard X-ray (14-195~KeV) sky are publicly available at  https://swift.gsfc.nasa.gov/results/bs157mon/671. 
In addition to the 8-band (14-20, 20-24, 24-35, 
35-50, 50-75, 75-100, 100-150, and 150-195~KeV) monthly averaged data, the website also provides 8-band 
snapshot light curves, starting from 2005. The snapshot light curves are extremely useful for exploring the short-timescale 
variability. The snapshot data is binned to generate 10-day binned light curves in different energy bands, 
then a total count rate for the 14-100~KeV energy  range. While binning, we flagged the 
low-exposure ($<$1~day) epochs to  reduce systematic errors. Given the low single-to-noise (S/N) ratio 
in bands 7 and 8, we discarded the 100-195~KeV energy band data.

Figure \ref{fig1} (a) shows the monthly averaged  hard X-ray (14-195~KeV) light curve from Dec.\ 2004 
until Dec.\ 2017 (blue circles). Prominent flux variations were detected in the source during 
this period.  The red squares show the  10-days binned light curves in the 14-100~KeV energy range. 
The count rates are scaled by a factor of 15 for visualization only. 
Given the high single-to-noise (S/N) ratio in the monthly binned data, the intensity variations can be studied 
in different energy bands. Photon flux light curves in different energy bands, 14-20, 20-24, 24-35, 
35-50, 50-75, 75-100, and 100-150~KeV, are plotted in panel (b). Band 8 (150-195~KeV) is not included in the plot because of low S/N  ratio. 
Similar variations are seen across the multiple bands. Variability is less pronounced at higher 
energy bands ($>$75~KeV) because of the low S/N ratio.

\begin{figure}
\includegraphics[scale=0.32, trim=0 1700 0 -1, clip]{Xray_flx_indx.eps}
\includegraphics[scale=0.38]{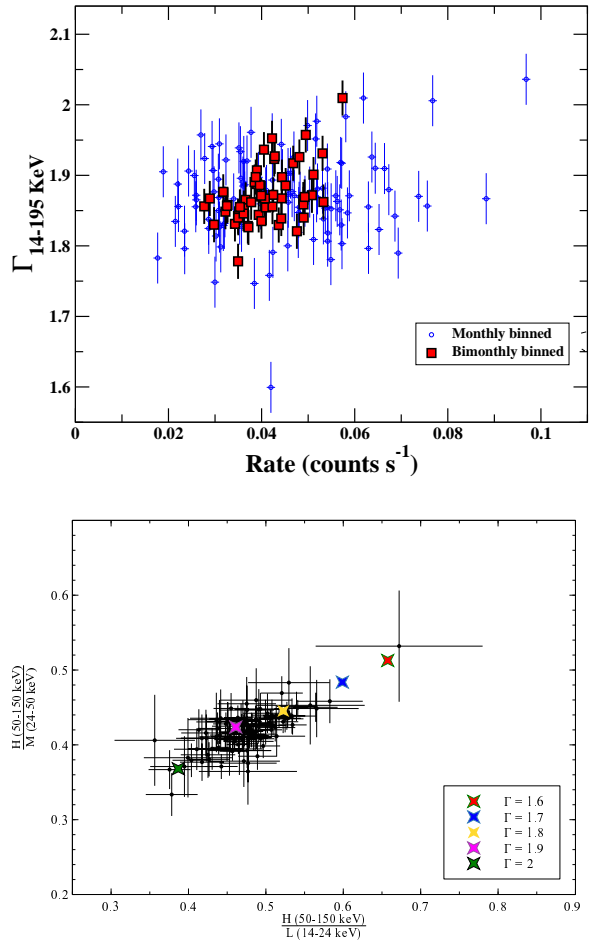}
\caption{{\it Top}: Photon index versus photon flux plot. {\it Bottom}: Spectral variability of Cen~A in the HR(hardness ratio)-plane. Black circles are the estimated hardness ratios in different energy channels, while the colored stars mark the simulated points. }
\label{fig2}
\end{figure}

Monthly (blue circles) and bimonthly (red squares) averaged hard X-ray photon index curve 
are illustrated in Fig.~\ref{fig1} (c). 
Despite significant flux variations in the X-ray bands, changes in the photon index  were 
small  but significant ($\Gamma$ $\sim$1.7 -- 1.9) until the end of 2012. 
Spectral variations  were more pronounced afterward.  Steepening of  the spectrum was observed 
until December 2013, reaching $\Gamma$ $\sim$2.0. Later, spectral hardening occurred until 
February 2015 ($\Gamma$ $\sim$1.7). The spectra soften back to the average value ($\Gamma$ $\sim$1.8) in recent years, with some mild variations.  In short, the hard X-ray spectra follow a steeper-when-brighter trend. The trend is evident in index versus photon flux plot (Fig.~\ref{fig2} top). A  linear Pearson correlation test is used
to quantify the correlation; we obtained $r_P$ (correlation coefficient) = 0.43 at the
99.8$\%$ confidence level for the bimonthly binned data, and $r_P$ = 0.16 at the
90$\%$ confidence level for the monthly binned data.

\begin{figure}
\includegraphics[scale=0.28, trim=0 1730 0 5, clip]{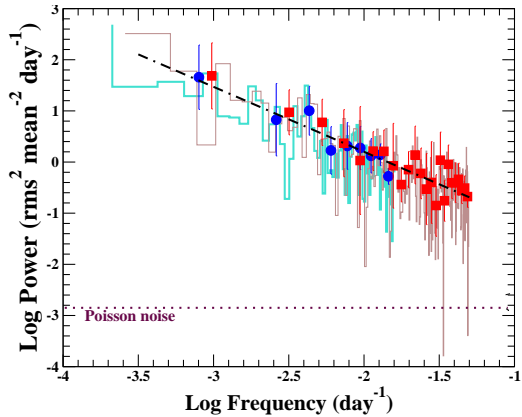}
\caption{Raw PSD of the monthly (in cyan) and 10-day (in brown) binned X-ray light curves. The dotted line 
marks the estimated Poisson noise level.  
The blue circles and red squares are the logarithmically binned PSDs for the monthly and 10-day binned 
data sets, respectively.  The black dashed line is the best fit power with a slope = 1.27$\pm$0.13. }
\label{fig_psd}
\end{figure}

We further investigate the spectral variability by analyzing hardness ratio (HR) time series. The monthly averaged data is used to calculate the time series in the following three energy bands: low or L-channel from 14-24 keV, medium or M-channel from 24-50 keV, and high or H-channel from 50-150 keV. We rebin the hard X-ray spectrum in this way in order to maximize the signal-to-noise.  (We exclude the 150-195 keV data due to low S/N ratio.) We then calculate HR values using those three channels, and confirm through a chi-squared test that the HR time series show statistically significant variability (p-value $<$0.05). Ratios of the time-series in different channels (HM versus HL) are then plotted against each other to produce an “HR-plane” (Fig.~\ref{fig2} black circles), and are used to investigate the nature of the spectral variability in the source.
A visual correlation can be seen 
between the HL and ML ratios. Linear Pearson correlation analysis confirms the correlation 
between the two. Formally we get, the Pearson correlation coefficient 
equals to 0.71 at $>$99.99$\%$ confidence level.
We further tested the spectral variations using simulations.  
The spectral simulations,  using different 
power law slopes, were performed using the '{\it fakeit}' command on Xspec \citep{arnaud1996}. 
The simulations were based on the 
simple "pegpwrlw" model, with $\Gamma$ in the range of 1 to 3 (typical photon indices of AGN in the 
{\it Swift}/BAT catalog). We then calculated the hardness ratios from the fake spectra (shown as 
colored stars in Fig.\ \ref{fig2}). 
The details of the simulations can be found in Mundo et al.\ 2022 (in preparation). 
The simulated points agree with the data, suggesting that the changing spectrum can be well described
as a simple power-law with a varying photon index over monthly timescales, spanning the range ~1.6--2.

\begin{figure*}
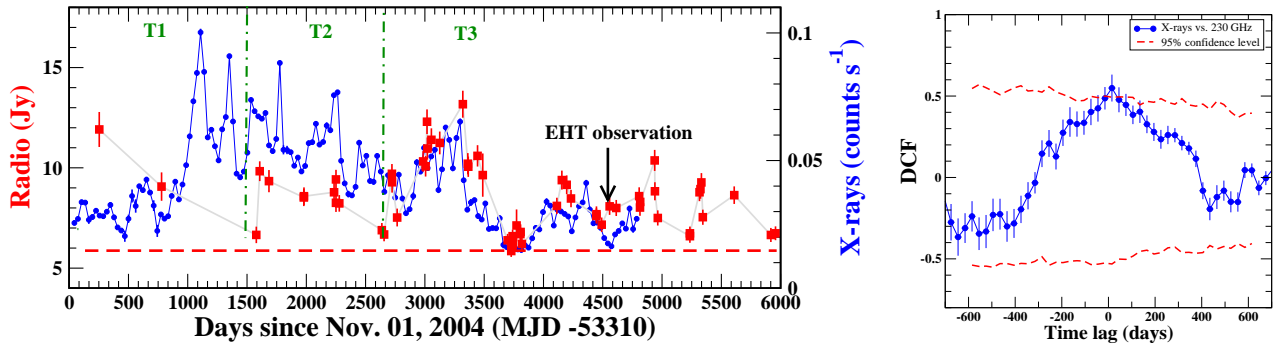
 
\includegraphics[scale=0.44, trim=0 0 0 -1, clip]{Xray_rad_plot.eps}
\hspace{8pt}
\includegraphics[scale=0.25, trim=0 0 0 -1, clip]{DCF.eps}
\caption{{\it [Left]} 230~GHz flux density light curve (red squares) superimposed 
on top of the 14-150~KeV light curve (blue circles). {\it [Right]} Cross-correlation analysis results: the DCF  curve is in blue, while the red dashed-lines mark the 95$\%$ confidence level. }
\label{fig3}
\end{figure*}

The nature of hard X-ray variability in the source is explored using the 
power spectrum density \citep[PSD][]{vaughan2003} analysis method. 
Both monthly averaged 14-195~KeV (data~A) and 10-day 
binned 14-100~KeV (data~B) light curves are used for the PSD analysis.
As a first step, we calculated the raw PSDs, squared modulus of the discrete Fourier transform. The raw PSDs are then logarithmically binned  to extract the 
slope of the underlying power spectrum, $P(f) \propto f^{\alpha}$ 
\citep[details are referred to][]{chidiac2016}. The Poisson noise level in the PSD is calculated following \citet{vaughan2003}.  The PSD analysis results are shown in 
Fig.~\ref{fig_psd}, where cyan and brown steps are the raw PSDs respectively for data~A and data~B, blue circles and red squares are their logarithmic binned values; errors mark the scatter of the raw PSD 
points. The best-fit power-law slope for data~A is $-(1.36\pm0.16)$, and for data~B is $-(1.29\pm0.11)$. A combined power-law fit  gives, $\alpha$ = $-(1.27\pm0.13)$ (black dashed 
line in Fig.~\ref{fig_psd}). We do not find any evidence of a break in the PSD, as is sometimes seen in the 2-10~KeV PSD of Seyfert galaxies \citep{vaughan2005a}.
A similar value of the PSD slope was reported by \citet{shimizu2013} using 58~months of 
the {\it Swift}/BAT data, and comparable  slopes have been seen in PSDs of beamed AGN  \citep{chidiac2016, algaba2018} as well. This 
implies that the hard X-ray variability of the source can be characterized simply as a red-noise process. 
Since there is no excess power at any frequency in the given time range, the PSD analysis rules out the presence of periodic variations  in the source.

\subsection{Radio SMA}
We used the 230 GHz data  provided by the Submillimeter Array (SMA) Observer Center\footnote{http://sma1.sma.hawaii.edu/callist/callist.html} database \citep{gurwell2007}
to investigate the flux variations in the radio regime. Figure \ref{fig3} (red squares) shows the flux density
variations observed in the source since July 2005.
The radio flux variations are superimposed on
top of a constant flux level of about 6~Jy (dashed line), and this could be related to the extended jet emission (see Section 3.1 for details).
Compared to X-rays (blue circles), the radio
data is sparsely sampled, especially in the beginning (segment T1). Some similarities in the long-term
decay trend can be seen in the two data sets over segment T2, and the flux variations are quite similar
afterward (segment T3).

\subsection{Cross-correlation}
The apparent correlation among X-ray and radio data sets was quantified using the discrete correlation 
function \citep[DCF][]{Edelson1988} method, and the significance of the correlation was 
tested  via simulations as discussed in \citet[Section A][]{rani2014}. The DCF results are 
presented in Fig.~\ref{fig3} (right). 
Monthly binned DCF points are in blue, while the red curves shows the 95$\%$ confidence levels. 
The DCF analysis of the two data sets shows a peak  above 95$\%$ confidence level at 0~days. 
This implies that the flux variations at X-ray and  regimes are well correlated with no 
time lag. Since the X-ray data is monthly sampled, a time-delay shorter than 30~days cannot be 
tested.  The correlation analysis therefore suggests that the hard X-rays and radio emission regions are co-spatial. This agrees with a tight correlation found among the parsec-scale radio luminosity and  X-ray luminosity of BAT detected Seyfert galaxies \citep{baek2019}. Radio and X-ray (especially soft X-rays) correlations have also been reported in several other Seyfert galaxies \citep{chatterjee2009, chatterjee2011, marscher2018} and explicitly used to probe the disk-jet connection in AGN.

%---------------------------
\section{Discussion}
Detailed spectral analysis \citep{furst2016} suggests that the hard X-ray emitting site is close to the central engine, but could not disentangle the ADAF and jet contribution. The multi-wavelength variability analysis presented here allows us to do so. 
Using 13 years of Swift/BAT and 230~GHz data, we performed a detailed temporal and correlation analysis, which revealed the followings. 
Prominent flux variations were observed in the source since 2004, but spectral changes were  rather moderate until 2012. Significant spectral variations were observed afterward, 
following a softer-when-brighter trend. 
The hard X-ray flux variability of the source is consistent with red-noise processes, with a slope $\sim -$1.3. 
Variations in the hard X-ray and 230~GHz radio data are  correlated, with no time lag. 
In the following subsections, we discuss the origin of hard X-ray emission in the context of ADAF and jet models.

\subsection{Nature of X-ray variability}
There have been many PSD studies of AGN, characterizing the PSD slopes, breaks, and their relation to the physical properties of the central engine. Breaks are a common feature in the PSDs of  Seyfert galaxies \citep{papadakis2004, done2005, markowitz2003}. These studies, however, are focused on the soft X-ray emission ($<$10~KeV), and the hard X-ray variability studies differ from this picture \citep{shimizu2013}.  The hard X-ray PSD of Cen~A is well-fitted using a slope of $-1.3$, with no evidence of a break. If we scale the breaks seen in the X-ray PSDs by the mass as in \citet{mchardy2004}, the predicted break timescale is higher than 20~days (log10~(Frequency) $>-1.3$~day$^{-1}$) and thus is not sensed by the BAT data. 
It is important to note that, unlike Seyfert galaxies, Cen~A is a low-luminosity radio-loud AGN, and as our study suggests, the PSD slope is comparable to that has been seen in PSDs of beamed AGN.

Except for the power-law slope ($\Gamma$ = 1.6--2.0), the X-ray spectrum of the 
source \citep[no reflection and very high cutoff-Energy,][]{furst2016} differs from that of Seyfert galaxies. Since the power-law slope from the jet and from the thermal Compotonization is very similar, the spectral slope cannot be used as a distinction between the two. However, the difference in source spectrum and temporal variations favor that the nature of X-ray emission in Cen~A is similar to beamed AGN.

\subsection{Nature of 230~GHz variability}
The  radio variations comprise of two components, quiescent and variable. 
As shown in Fig.\ \ref{fig3}, even at 230~GHz, we have about 6~Jy 
contribution from the extended jet region. Earlier studies found a  
contribution of about 7~Jy from the extended jet emission in the total flux density of the source \citep{israel2008}. 
Since Cen~A has a complex extended jet structure, the location of the variable component remained unclear.  
Using continuum observations in the millimeter and sub-millimeter regime, \citet{israel2008} investigated 
the flux and spectral variability of the source, and reported that most (if not all)  variations are from the 
milli-arcsecond core. On micro-arcsecond scales, the Event Horizon Telescope (EHT) discovered a completely 
different picture of the core \citep{eht2021}. The core of the source is opaque at 230~GHz and the 
turnover frequency  is at $\sim$THz frequencies. The source has a flux density of $\sim$2~Jy with an 
edge-brightened jet.  It is quite probable that either the flux density of the core or of the two 
lanes is varying, but  multiple observations are required to confirm that. The radio luminosity of the 
source, measured by the EHT, is 7.5$\times$10$^{39}$~erg~s$^{-1}$. However, EHT observed the source not in its 
brightest phase (see Fig.~\ref{fig3}). After subtracting the quiescent flux (6~Jy) from the total flux, the peak flux of the variable component is $\sim$7~Jy, which corresponds to  2.6$\times$10$^{40}$~erg~s$^{-1}$.

\subsection{ADAF versus jet models}
In an ADAF model, the radio emission is because of cyclo-synchrotron radiation from hot electrons in the
equipartition magnetic field; it should be isotropic. In the absence of a radio jet, the expected radio luminosity is
roughly proportional to the mass of the central black hole and its accretion rate \citep{yi1999, mahadaven1997}, and
is given by
\begin{equation}
L_{230~GHz, ADAF}\sim 2.5\times 10^{38} m_7^{8/5} {\dot m}_{-3}^{6/5}~erg/s
\end{equation}
where
$m_7$ is the black hole mass in units of 10$^{7}$~M$_{\odot}$ and ${\dot m}_{-3}$ = ${\dot m}/10^{-3}$,  ${\dot m}$
is accretion in units of Eddington rate. Using $m_7$ = 5 \citep{neumayer2010} and ${\dot m}_{-3}$ = 0.2 \citep{evans2004},
the 230~GHz ADAF luminosity for Cen~A is $\sim$5$\times$10$^{38}$~erg~s$^{-1}$, which is significantly
lower than the observed 230~GHz radio luminosity.  This implies that the  ADAF component has a negligible
contribution to the observed radio luminosity of the source.

Jet luminosity, in case of mainly powered by black hole accretion,  can be estimated using eq.~9 in \citet{eht2021},
\begin{multline*}
%\begin{equation}
    P_{\rm jet}=2.2\times10^{43}f(a_{*})\left(\frac{\phi}{15}\right)^{2}
    \left(\frac{\dot{M}}{10^{-6}\dot{M}_{\rm Edd}}\right) \\
    \left( \frac{M}{6.2\times10^{9}M_{\odot}}\right) \rm erg\,s^{-1}
%\end{equation}
\end{multline*}
where
$0\leq a_{*} \leq 1$ is the normalized black hole spin,
$1\leq \phi \leq 15$ is the normalized magnetic flux at the black hole event horizon,
$f(a_{*})\approx a_{*}^{2}\left(1+\sqrt{(1-a_{*}^{2})}\right)^{-2}$ (for $a_{*}<0.95$), 
$\dot{M}=2\times10^{-4}\dot{M}_{\rm edd}$, and
$M=5\times10^{7}M_{\odot}$.
For $a_{*}\leq0.2$ and $\phi\leq1$, we have a marginally low jet power of $P_{\rm jet}\leq \times10^{39}\rm erg\,s^{-1}$. 
Slightly larger values of $a_{*}=0.3$ and $\phi=2$ gives $P_{\rm jet}$ $\sim 1.5\times10^{40}\rm erg\,s^{-1}$, which well explains the observed radio luminosity.

Both ADAF and jet models predict a strong correlation between radio and X-ray flux variations. However, the
ADAF models predict a very characteristic spectrum of slope 1/3 in the radio regime \citep{mahadaven1997}. The observed radio spectrum of the
source has a slope of $\sim$ 0.7 below the turnover frequency (around 5 to 20~GHz). A slightly steeper spectrum, slope ~0.8,
is observed at higher frequencies \citep{israel2008}. This implies that the observed radio spectrum of the source
is not consistent with the ADAF model.  Moreover, the shape of the hard X-ray spectrum in ADAF models is thermal bremsstrahlung not
a power-law. For low luminosity AGN \citep[accreting close to m$_{critical}$ = 0.003 to 0.02,][]{mahadaven1997, narayan1996}, X-ray spectrum is very hard ($\Gamma$ $\sim$0.7). The power-law spectrum of Cen~A, $\Gamma$ varying between 1.7 to 2, rules out the ADAF models.
Another factor that argues in favor of the    jet-based origin of the hard X-ray
emission is the steeping of the X-ray spectrum as the source gets brighter. ADAF models predict harder-when-brighter behavior \citep{esin1997}. As discussed in Section 2.1, the spectrum gets steeper while the source gets brighter.

\section{Conclusions}
We present a thorough  analysis of hard X-ray emission from Cen~A using 13 years (Dec.~2004--Dec.~2017) of {\it Swift}/BAT observations. Prominent photon flux variations were detected during this period, and the variations   are consistent with a red-noise process of slope $-1.3$. The source spectral variations were  rather moderate until the end of 2012; a steeper-while-brighter trend was observed afterward. We found a significant correlation between the hard X-ray and 230~GHz flux variations with no time-lag, indicating a co-spatiality of their emitting sites.

Previous spectral  analysis confirms that hard X-ray emission of the source is confined within the 
core, and is produced either via Comptonization in an ADAF flow or at the base of the inner jet \citep{furst2016}. The study could not disentangle the two. 
However, variability analysis and the broadband spectral energy distribution studies of the source, using decade long {\it Rossi X-ray Timing Explorer (RXTE)} observations, favored the jet based origin of the hard X-ray emission \citep{rothschild2011}.
Using a comprehensive analysis of the hard X-ray emission and its correlation with the 230~GHz observations, we probe the  hard X-ray emitting site in Cen~A. The
following arguments rule out the ADAF models: (1) observed 230~GHz luminosity is significantly higher than L$_{230~GHz, ADAF}$, (2)  radio spectral slope ($\sim$0.7-0.8) contradicts with the characteristic slope of 1/3 predicted by the ADAF models, (3) power-law X-ray spectral shape, and (4) a softer-when brighter behavior of the hard X-ray spectra. The study confirms the jet-based origin of the hard X-ray emission in the source.

%%%%%%%%%%%%%%%%%%%%%%%%%%%%%%%%%%%%%%%%%%%%%%%%%%%%%%%%%%%%%%%%%%%%
\begin{acknowledgments}
BR acknowledges support from the National Research Foundation of Korea (grant number 2021R1A2C1095799). 
JYK acknowledges support from the National Research Foundation of Korea (grant number 2022R1C1C1005255). The Submillimeter Array is a joint project between the Smithsonian Astrophysical Observatory and the Academia Sinica Institute of Astronomy and Astrophysics and is funded by the Smithsonian Institution and the Academia Sinica. We thank the reviewer for an insightful review of the paper.
\end{acknowledgments}

%%%%%%%%%%%%%%%%%%%%%%%%%%%%%%%%%%%%%%%%%%%%%%%%%%%%%%%%%%%%%%%%%%%%
%%%%%%%%%%%%%%%%%%%%%%%%%%%%%%%%%%%%%%%%%%%%%%%%%%%%%%%%%%%%%%%%%%%%

%\bibliographystyle{aasjournal}
%\bibliography{references}

\begin{thebibliography}{}
\expandafter\ifx\csname natexlab\endcsname\relax\def\natexlab#1{#1}\fi
\providecommand{\url}[1]{\href{#1}{#1}}
\providecommand{\dodoi}[1]{doi:~\href{http://doi.org/#1}{\nolinkurl{#1}}}
\providecommand{\doeprint}[1]{\href{http://ascl.net/#1}{\nolinkurl{http://ascl.net/#1}}}
\providecommand{\doarXiv}[1]{\href{https://arxiv.org/abs/#1}{\nolinkurl{https://arxiv.org/abs/#1}}}

\bibitem[{{Abdo} {et~al.}(2010){Abdo}, {Ackermann}, {Ajello}, {Atwood},
  {Baldini}, {Ballet}, {Barbiellini}, {Bastieri}, {Baughman}, {Bechtol},
  {Bellazzini}, {Berenji}, {Blandford}, {Bloom}, {Bonamente}, {Borgland},
  {Bregeon}, {Brez}, {Brigida}, {Bruel}, {Burnett}, {Buson}, {Caliandro},
  {Cameron}, {Caraveo}, {Casandjian}, {Cavazzuti}, {Cecchi}, {{\c{C}}elik},
  {Chekhtman}, {Cheung}, {Chiang}, {Ciprini}, {Claus}, {Cohen-Tanugi},
  {Colafrancesco}, {Cominsky}, {Conrad}, {Costamante}, {Cutini}, {Davis},
  {Dermer}, {de Angelis}, {de Palma}, {Digel}, {do Couto e Silva}, {Drell},
  {Dubois}, {Dumora}, {Farnier}, {Favuzzi}, {Fegan}, {Finke}, {Focke},
  {Fortin}, {Fukazawa}, {Funk}, {Fusco}, {Gargano}, {Gasparrini}, {Gehrels},
  {Georganopoulos}, {Germani}, {Giebels}, {Giglietto}, {Giordano}, {Giroletti},
  {Glanzman}, {Godfrey}, {Grenier}, {Grove}, {Guillemot}, {Guiriec},
  {Hanabata}, {Harding}, {Hayashida}, {Hays}, {Hughes}, {Jackson},
  {J{\'o}hannesson G.}, {Johnson}, {Johnson}, {Johnson}, {Kamae}, {Katagiri},
  {Kataoka}, {Kawai}, {Kerr}, {Kn{\"o}dlseder}, {Kocian}, {Kuss}, {Lande},
  {Latronico}, {Lemoine-Goumard}, {Longo}, {Loparco}, {Lott}, {Lovellette},
  {Lubrano}, {Madejski}, {Makeev}, {Mazziotta}, {McConville}, {McEnery},
  {Meurer}, {Michelson}, {Mitthumsiri}, {Mizuno}, {Moiseev}, {Monte},
  {Monzani}, {Morselli}, {Moskalenko}, {Murgia}, {Nolan}, {Norris}, {Nuss},
  {Ohsugi}, {Omodei}, {Orlando}, {Ormes}, {Paneque}, {Parent}, {Pelassa},
  {Pepe}, {Pesce-Rollins}, {Piron}, {Porter}, {Rain{\`o}}, {Rando}, {Razzano},
  {Razzaque}, {Reimer}, {Reimer}, {Reposeur}, {Ritz}, {Rochester}, {Rodriguez},
  {Romani}, {Roth}, {Ryde}, {Sadrozinski}, {Sambruna}, {Sanchez}, {Sander},
  {Saz Parkinson}, {Scargle}, {Sgr{\`o}}, {Siskind}, {Smith}, {Smith},
  {Spandre}, {Spinelli}, {Starck}, {Stawarz}, {Strickman}, {Suson}, {Tajima},
  {Takahashi}, {Takahashi}, {Tanaka}, {Thayer}, {Thayer}, {Thompson},
  {Tibaldo}, {Torres}, {Tosti}, {Tramacere}, {Uchiyama}, {Vasileiou},
  {Vilchez}, {Vitale}, {Waite}, {Wallace}, {Wang}, {Winer}, {Wood}, {Ylinen},
  {Ziegler}, {Hardcastle}, {Kazanas}, \& {Fermi LAT Collaboration}}]{abdo2010}
{Abdo}, A.~A., {Ackermann}, M., {Ajello}, M., {et~al.} 2010, Science, 328, 725,
  \dodoi{10.1126/science.1184656}

\bibitem[{{Aharonian} {et~al.}(2009){Aharonian}, {Akhperjanian}, {Anton}, {de
  Almeida}, {Bazer-Bachi}, {Becherini}, {Behera}, {Benbow}, {Bernl{\"o}hr},
  {Boisson}, {Bochow}, {Borrel}, {Brion}, {Brucker}, {Brun}, {B{\"u}hler},
  {Bulik}, {B{\"u}sching}, {Boutelier}, {Chadwick}, {Charbonnier}, {Chaves},
  {Cheesebrough}, {Chounet}, {Clapson}, {Coignet}, {Dalton}, {Daniel},
  {Davids}, {Degrange}, {Deil}, {Dickinson}, {Djannati-Ata{\"\i}}, {Domainko},
  {Drury}, {Dubois}, {Dubus}, {Dyks}, {Dyrda}, {Egberts}, {Emmanoulopoulos},
  {Espigat}, {Farnier}, {Feinstein}, {Fiasson}, {F{\"o}rster}, {Fontaine},
  {F{\"u}{\ss}ling}, {Gabici}, {Gallant}, {G{\'e}rard}, {Giebels},
  {Glicenstein}, {Gl{\"u}ck}, {Goret}, {G{\"o}hring}, {Hauser}, {Hauser},
  {Heinz}, {Heinzelmann}, {Henri}, {Hermann}, {Hinton}, {Hoffmann}, {Hofmann},
  {Holleran}, {Hoppe}, {Horns}, {Jacholkowska}, {de Jager}, {Jahn}, {Jung},
  {Katarzy{\'n}ski}, {Katz}, {Kaufmann}, {Kendziorra}, {Kerschhaggl},
  {Khangulyan}, {Kh{\'e}lifi}, {Keogh}, {Klu{\'z}niak}, {Kneiske}, {Komin},
  {Kosack}, {Lamanna}, {Latham}, {Lenain}, {Lohse}, {Marandon}, {Martin},
  {Martineau-Huynh}, {Marcowith}, {Maurin}, {McComb}, {Medina}, {Moderski},
  {Moulin}, {Naumann-Godo}, {de Naurois}, {Nedbal}, {Nekrassov}, {Niemiec},
  {Nolan}, {Ohm}, {Olive}, {de O{\~n}a Wilhelmi}, {Orford}, {Ostrowski},
  {Panter}, {Arribas}, {Pedaletti}, {Pelletier}, {Petrucci}, {Pita},
  {P{\"u}hlhofer}, {Punch}, {Quirrenbach}, {Raubenheimer}, {Raue}, {Rayner},
  {Renaud}, {Rieger}, {Ripken}, {Rob}, {Rosier-Lees}, {Rowell}, {Rudak},
  {Rulten}, {Ruppel}, {Sahakian}, {Santangelo}, {Schlickeiser}, {Sch{\"o}ck},
  {Schr{\"o}der}, {Schwanke}, {Schwarzburg}, {Schwemmer}, {Shalchi}, {Sikora},
  {Skilton}, {Sol}, {Spangler}, {Stawarz}, {Steenkamp}, {Stegmann}, {Superina},
  {Szostek}, {Tam}, {Tavernet}, {Terrier}, {Tibolla}, {Tluczykont}, {van
  Eldik}, {Vasileiadis}, {Venter}, {Venter}, {Vialle}, {Vincent}, {Vink},
  {Vivier}, {V{\"o}lk}, {Volpe}, {Wagner}, {Ward}, {Zdziarski}, \&
  {Zech}}]{hess2009}
{Aharonian}, F., {Akhperjanian}, A.~G., {Anton}, G., {et~al.} 2009, \apjl, 695,
  L40, \dodoi{10.1088/0004-637X/695/1/L40}

\bibitem[{{Algaba} {et~al.}(2018){Algaba}, {Lee}, {Kim}, {Rani}, {Hodgson},
  {Kino}, {Trippe}, {Park}, {Zhao}, {Byun}, {Gurwell}, {Kang}, {Kim}, {Kim},
  {Kim}, {Lott}, {Miyazaki}, \& {Wajima}}]{algaba2018}
{Algaba}, J.-C., {Lee}, S.-S., {Kim}, D.-W., {et~al.} 2018, \apj, 852, 30,
  \dodoi{10.3847/1538-4357/aa9e50}

\bibitem[{{Arnaud}(1996)}]{arnaud1996}
{Arnaud}, K.~A. 1996, in Astronomical Society of the Pacific Conference Series,
  Vol. 101, Astronomical Data Analysis Software and Systems V, ed. G.~H.
  {Jacoby} \& J.~{Barnes}, 17

\bibitem[{{Baek} {et~al.}(2019){Baek}, {Chung}, {Schawinski}, {Oh}, {Wong},
  {Koss}, {Ricci}, {Trakhtenbrot}, {Smith}, \& {Ueda}}]{baek2019}
{Baek}, J., {Chung}, A., {Schawinski}, K., {et~al.} 2019, \mnras, 488, 4317,
  \dodoi{10.1093/mnras/stz1995}

\bibitem[{{Baity} {et~al.}(1981){Baity}, {Rothschild}, {Lingenfelter}, {Stein},
  {Nolan}, {Gruber}, {Knight}, {Matteson}, {Peterson}, {Primini}, {Levine},
  {Lewin}, {Mushotzky}, \& {Tennant}}]{baity1981}
{Baity}, W.~A., {Rothschild}, R.~E., {Lingenfelter}, R.~E., {et~al.} 1981,
  \apj, 244, 429, \dodoi{10.1086/158719}

\bibitem[{{Beckmann} {et~al.}(2011){Beckmann}, {Jean}, {Lubi{\'n}ski}, {Soldi},
  \& {Terrier}}]{beckmann2011}
{Beckmann}, V., {Jean}, P., {Lubi{\'n}ski}, P., {Soldi}, S., \& {Terrier}, R.
  2011, \aap, 531, A70, \dodoi{10.1051/0004-6361/201016020}

\bibitem[{{Burke} {et~al.}(2014){Burke}, {Jourdain}, {Roques}, \&
  {Evans}}]{burke2014}
{Burke}, M.~J., {Jourdain}, E., {Roques}, J.-P., \& {Evans}, D.~A. 2014, \apj,
  787, 50, \dodoi{10.1088/0004-637X/787/1/50}

\bibitem[{{Chatterjee} {et~al.}(2009){Chatterjee}, {Marscher}, {Jorstad},
  {Olmstead}, {McHardy}, {Aller}, {Aller}, {L{\"a}hteenm{\"a}ki}, {Tornikoski},
  {Hovatta}, {Marshall}, {Miller}, {Ryle}, {Chicka}, {Benker}, {Bottorff},
  {Brokofsky}, {Campbell}, {Chonis}, {Gaskell}, {Gaynullina}, {Grankin},
  {Hedrick}, {Ibrahimov}, {Klimek}, {Kruse}, {Masatoshi}, {Miller}, {Pan},
  {Petersen}, {Peterson}, {Shen}, {Strel'nikov}, {Tao}, {Watkins}, \&
  {Wheeler}}]{chatterjee2009}
{Chatterjee}, R., {Marscher}, A.~P., {Jorstad}, S.~G., {et~al.} 2009, \apj,
  704, 1689, \dodoi{10.1088/0004-637X/704/2/1689}

\bibitem[{{Chatterjee} {et~al.}(2011){Chatterjee}, {Marscher}, {Jorstad},
  {Markowitz}, {Rivers}, {Rothschild}, {McHardy}, {Aller}, {Aller},
  {L{\"a}hteenm{\"a}ki}, {Tornikoski}, {Harrison}, {Agudo}, {G{\'o}mez},
  {Taylor}, \& {Gurwell}}]{chatterjee2011}
---. 2011, \apj, 734, 43, \dodoi{10.1088/0004-637X/734/1/43}

\bibitem[{{Chidiac} {et~al.}(2016){Chidiac}, {Rani}, {Krichbaum}, {Angelakis},
  {Fuhrmann}, {Nestoras}, {Zensus}, {Sievers}, {Ungerechts}, {Itoh},
  {Fukazawa}, {Uemura}, {Sasada}, {Gurwell}, \& {Fedorova}}]{chidiac2016}
{Chidiac}, C., {Rani}, B., {Krichbaum}, T.~P., {et~al.} 2016, \aap, 590, A61,
  \dodoi{10.1051/0004-6361/201628347}

\bibitem[{{Done} \& {Gierli{\'n}ski}(2005)}]{done2005}
{Done}, C., \& {Gierli{\'n}ski}, M. 2005, \mnras, 364, 208,
  \dodoi{10.1111/j.1365-2966.2005.09555.x}

\bibitem[{{Edelson} \& {Krolik}(1988)}]{Edelson1988}
{Edelson}, R.~A., \& {Krolik}, J.~H. 1988, \apj, 333, 646,
  \dodoi{10.1086/166773}

\bibitem[{{Esin} {et~al.}(1997){Esin}, {McClintock}, \& {Narayan}}]{esin1997}
{Esin}, A.~A., {McClintock}, J.~E., \& {Narayan}, R. 1997, \apj, 489, 865,
  \dodoi{10.1086/304829}

\bibitem[{{Evans} {et~al.}(2004){Evans}, {Kraft}, {Worrall}, {Hardcastle},
  {Jones}, {Forman}, \& {Murray}}]{evans2004}
{Evans}, D.~A., {Kraft}, R.~P., {Worrall}, D.~M., {et~al.} 2004, \apj, 612,
  786, \dodoi{10.1086/422806}

\bibitem[{{Fanaroff} \& {Riley}(1974)}]{fanaroff74}
{Fanaroff}, B.~L., \& {Riley}, J.~M. 1974, \mnras, 167, 31P,
  \dodoi{10.1093/mnras/167.1.31P}

\bibitem[{{Fukazawa} {et~al.}(2011){Fukazawa}, {Hiragi}, {Yamazaki}, {Mizuno},
  {Hayashi}, {Hayashi}, {Nishino}, {Takahashi}, \& {Ohno}}]{fukawaza2011}
{Fukazawa}, Y., {Hiragi}, K., {Yamazaki}, S., {et~al.} 2011, \apj, 743, 124,
  \dodoi{10.1088/0004-637X/743/2/124}

\bibitem[{{F{\"u}rst} {et~al.}(2016){F{\"u}rst}, {M{\"u}ller}, {Madsen},
  {Lanz}, {Rivers}, {Brightman}, {Arevalo}, {Balokovi{\'c}}, {Beuchert},
  {Boggs}, {Christensen}, {Craig}, {Dauser}, {Farrah}, {Graefe}, {Hailey},
  {Harrison}, {Kadler}, {King}, {Krau{\ss}}, {Madejski}, {Matt}, {Marinucci},
  {Markowitz}, {Ogle}, {Ojha}, {Rothschild}, {Stern}, {Walton}, {Wilms}, \&
  {Zhang}}]{furst2016}
{F{\"u}rst}, F., {M{\"u}ller}, C., {Madsen}, K.~K., {et~al.} 2016, \apj, 819,
  150, \dodoi{10.3847/0004-637X/819/2/150}

\bibitem[{{Gurwell} {et~al.}(2007){Gurwell}, {Peck}, {Hostler}, {Darrah}, \&
  {Katz}}]{gurwell2007}
{Gurwell}, M.~A., {Peck}, A.~B., {Hostler}, S.~R., {Darrah}, M.~R., \& {Katz},
  C.~A. 2007, in Astronomical Society of the Pacific Conference Series, Vol.
  375, From Z-Machines to ALMA: (Sub)Millimeter Spectroscopy of Galaxies, ed.
  A.~J. {Baker}, J.~{Glenn}, A.~I. {Harris}, J.~G. {Mangum}, \& M.~S. {Yun},
  234

\bibitem[{{H.~E.~S.~S. Collaboration} {et~al.}(2020){H.~E.~S.~S.
  Collaboration}, {Abdalla}, {Adam}, {Aharonian}, {Ait Benkhali},
  {Ang{\"u}ner}, {Arakawa}, {Arcaro}, {Armand}, {Ashkar}, {Backes}, {Barbosa
  Martins}, {Barnard}, {Becherini}, {Berge}, {Bernl{\"o}hr}, {Blackwell},
  {B{\"o}ttcher}, {Boisson}, {Bolmont}, {Bonnefoy}, {Bregeon}, {Breuhaus},
  {Brun}, {Brun}, {Bryan}, {B{\"u}chele}, {Bulik}, {Bylund}, {Capasso},
  {Caroff}, {Carosi}, {Casanova}, {Cerruti}, {Chand}, {Chandra}, {Chen},
  {Colafrancesco}, {Cury{\l}o}, {Davids}, {Deil}, {Devin}, {deWilt}, {Dirson},
  {Djannati-Ata{\"\i}}, {Dmytriiev}, {Donath}, {Doroshenko}, {Drury}, {Dyks},
  {Egberts}, {Emery}, {Ernenwein}, {Eschbach}, {Feijen}, {Fegan}, {Fiasson},
  {Fontaine}, {Funk}, {F{\"u}{\ss}ling}, {Gabici}, {Gallant}, {Gat{\'e}},
  {Giavitto}, {Glawion}, {Glicenstein}, {Gottschall}, {Grondin}, {Hahn},
  {Haupt}, {Heinzelmann}, {Henri}, {Hermann}, {Hinton}, {Hofmann}, {Hoischen},
  {Holch}, {Holler}, {Horns}, {Huber}, {Iwasaki}, {Jamrozy}, {Jankowsky},
  {Jankowsky}, {Jardin-Blicq}, {Jung-Richardt}, {Kastendieck},
  {Katarzy{\'n}ski}, {Katsuragawa}, {Katz}, {Khangulyan}, {Kh{\'e}lifi},
  {King}, {Klepser}, {Klu{\'z}niak}, {Komin}, {Kosack}, {Kostunin}, {Kraus},
  {Lamanna}, {Lau}, {Lemi{\`e}re}, {Lemoine-Goumard}, {Lenain}, {Leser},
  {Levy}, {Lohse}, {Lypova}, {Mackey}, {Majumdar}, {Malyshev}, {Marandon},
  {Marcowith}, {Mares}, {Mariaud}, {Mart{\'\i}-Devesa}, {Marx}, {Maurin},
  {Meintjes}, {Mitchell}, {Moderski}, {Mohamed}, {Mohrmann}, {Moore}, {Moulin},
  {Muller}, {Murach}, {Nakashima}, {de Naurois}, {Ndiyavala}, {Niederwanger},
  {Niemiec}, {Oakes}, {O'Brien}, {Odaka}, {Ohm}, {de Ona Wilhelmi},
  {Ostrowski}, {Oya}, {Panter}, {Parsons}, {Perennes}, {Petrucci}, {Peyaud},
  {Piel}, {Pita}, {Poireau}, {Priyana Noel}, {Prokhorov}, {Prokoph},
  {P{\"u}hlhofer}, {Punch}, {Quirrenbach}, {Raab}, {Rauth}, {Reimer}, {Reimer},
  {Remy}, {Renaud}, {Rieger}, {Rinchiuso}, {Romoli}, {Rowell}, {Rudak},
  {Ruiz-Velasco}, {Sahakian}, {Saito}, {Sanchez}, {Santangelo}, {Sasaki},
  {Schlickeiser}, {Sch{\"u}ssler}, {Schulz}, {Schutte}, {Schwanke},
  {Schwemmer}, {Seglar-Arroyo}, {Senniappan}, {Seyffert}, {Shafi},
  {Shiningayamwe}, {Simoni}, {Sinha}, {Sol}, {Specovius}, {Spir-Jacob},
  {Stawarz}, {Steenkamp}, {Stegmann}, {Steppa}, {Takahashi}, {Tavernier},
  {Taylor}, {Terrier}, {Tiziani}, {Tluczykont}, {Trichard}, {Tsirou}, {Tsuji},
  {Tuffs}, {Uchiyama}, {van der Walt}, {van Eldik}, {van Rensburg}, {van
  Soelen}, {Vasileiadis}, {Veh}, {Venter}, {Vincent}, {Vink}, {Voisin},
  {V{\"o}lk}, {Vuillaume}, {Wadiasingh}, {Wagner}, {White}, {Wierzcholska},
  {Yang}, {Yoneda}, {Zacharias}, {Zanin}, {Zdziarski}, {Zech}, {Ziegler},
  {Zorn}, \& {{\.Z}ywucka}}]{hess2020}
{H.~E.~S.~S. Collaboration}, {Abdalla}, H., {Adam}, R., {et~al.} 2020, \nat,
  582, 356, \dodoi{10.1038/s41586-020-2354-1}

\bibitem[{{Hardcastle} {et~al.}(2003){Hardcastle}, {Worrall}, {Kraft},
  {Forman}, {Jones}, \& {Murray}}]{hardcastle2003}
{Hardcastle}, M.~J., {Worrall}, D.~M., {Kraft}, R.~P., {et~al.} 2003, \apj,
  593, 169, \dodoi{10.1086/376519}

\bibitem[{{Harris} {et~al.}(2010){Harris}, {Rejkuba}, \& {Harris}}]{harris10}
{Harris}, G. L.~H., {Rejkuba}, M., \& {Harris}, W.~E. 2010, \pasa, 27, 457,
  \dodoi{10.1071/AS09061}

\bibitem[{{Hinkle} \& {Mushotzky}(2021)}]{hinkle2021}
{Hinkle}, J.~T., \& {Mushotzky}, R. 2021, \mnras, 506, 4960,
  \dodoi{10.1093/mnras/stab1976}

\bibitem[{{Israel} {et~al.}(2008){Israel}, {Raban}, {Booth}, \&
  {Rantakyr{\"o}}}]{israel2008}
{Israel}, F.~P., {Raban}, D., {Booth}, R.~S., \& {Rantakyr{\"o}}, F.~T. 2008,
  \aap, 483, 741, \dodoi{10.1051/0004-6361:20079229}

\bibitem[{{Janssen} {et~al.}(2021){Janssen}, {Falcke}, {Kadler}, {Ros},
  {Wielgus}, {Akiyama}, {Balokovi{\'c}}, {Blackburn}, {Bouman}, {Chael},
  {Chan}, {Chatterjee}, {Davelaar}, {Edwards}, {Fromm}, {G{\'o}mez}, {Goddi},
  {Issaoun}, {Johnson}, {Kim}, {Koay}, {Krichbaum}, {Liu}, {Liuzzo}, {Markoff},
  {Markowitz}, {Marrone}, {Mizuno}, {M{\"u}ller}, {Ni}, {Pesce},
  {Ramakrishnan}, {Roelofs}, {Rygl}, {van Bemmel}, {Event Horizon Telescope
  Collaboration}, {Alberdi}, {Alef}, {Algaba}, {Anantua}, {Asada}, {Azulay},
  {Baczko}, {Ball}, {Ball}, {Barrett}, {Benson}, {Bintley}, {Bintley},
  {Blundell}, {Boland}, {Boland}, {Bower}, {Boyce}, {Bremer}, {Brinkerink},
  {Brissenden}, {Britzen}, {Broderick}, {Broguiere}, {Bronzwaer}, {Byun},
  {Carlstrom}, {Chatterjee}, {Chen}, {Chen}, {Chesler}, {Cho}, {Christian},
  {Conway}, {Cordes}, {Crawford}, {Crew}, {Cruz-Osorio}, {Cui}, {Cui}, {De
  Laurentis}, {Deane}, {Dempsey}, {Desvignes}, {Dexter}, {Doeleman}, {Eatough},
  {Farah}, {Farah}, {Fish}, {Fomalont}, {Ford}, {Fraga-Encinas}, {Friberg},
  {Friberg}, {Fuentes}, {Galison}, {Gammie}, {Garc{\'\i}a}, {Gelles}, {Gentaz},
  {Georgiev}, {Georgiev}, {Gold}, {Gold}, {G{\'o}mez-Ruiz}, {Gu}, {Gurwell},
  {Hada}, {Haggard}, {Hecht}, {Hesper}, {Himwich}, {Ho}, {Ho}, {Honma},
  {Huang}, {Huang}, {Hughes}, {Ikeda}, {Inoue}, {Inoue}, {James}, {Jannuzi},
  {Jannuzi}, {Jeter}, {Jiang}, {Jimenez-Rosales}, {Jimenez-Rosales}, {Jorstad},
  {Jung}, {Karami}, {Karuppusamy}, {Kawashima}, {Keating}, {Kettenis}, {Kim},
  {Kim}, {Kim}, {Kim}, {Kino}, {Kino}, {Kofuji}, {Koyama}, {Kramer}, {Kramer},
  {Kramer}, {Kuo}, {Lauer}, {Lee}, {Levis}, {Li}, {Li}, {Lindqvist}, {Lico},
  {Lindahl}, {Lindahl}, {Liu}, {Liu}, {Lo}, {Lobanov}, {Loinard}, {Lonsdale},
  {Lu}, {MacDonald}, {Mao}, {Marchili}, {Marchili}, {Marchili}, {Marscher},
  {Mart{\'\i}-Vidal}, {Matsushita}, {Matthews}, {Medeiros}, {Menten}, {Mizuno},
  {Mizuno}, {Moran}, {Moriyama}, {Moscibrodzka}, {Moscibrodzka}, {Musoke},
  {Mej{\'\i}as}, {Nagai}, {Nagar}, {Nakamura}, {Narayan}, {Narayanan},
  {Natarajan}, {Nathanail}, {Neilsen}, {Neri}, {Neri}, {Noutsos}, {Nowak},
  {Okino}, {Olivares}, {Ortiz-Le{\'o}n}, {Oyama}, {{\"O}zel}, {Palumbo},
  {Park}, {Patel}, {Pen}, {Pen}, {Pi{\'e}tu}, {Plambeck}, {PopStefanija},
  {Porth}, {P{\"o}tzl}, {Prather}, {Preciado-L{\'o}pez}, {Psaltis}, {Pu}, {Pu},
  {Rao}, {Rawlings}, {Raymond}, {Rezzolla}, {Ricarte}, {Ripperda}, {Ripperda},
  {Rogers}, {Rogers}, {Rose}, {Roshanineshat}, {Rottmann}, {Roy}, {Ruszczyk},
  {Ruszczyk}, {S{\'a}nchez}, {S{\'a}nchez-Arguelles}, {Sasada}, {Savolainen},
  {Schloerb}, {Schuster}, {Shao}, {Shen}, {Small}, {Sohn}, {SooHoo}, {Sun},
  {Tazaki}, {Tetarenko}, {Tiede}, {Tilanus}, {Titus}, {Torne}, {Trent},
  {Traianou}, {Trippe}, {van Bemmel}, {van Langevelde}, {van Rossum}, {Wagner},
  {Ward-Thompson}, {Wardle}, {Weintroub}, {Wex}, {Wharton}, {Wharton}, {Wong},
  {Wu}, {Yoon}, {Young}, {Young}, {Younsi}, {Yuan}, {Yuan}, {Zensus}, {Zhao},
  \& {Zhao}}]{eht2021}
{Janssen}, M., {Falcke}, H., {Kadler}, M., {et~al.} 2021, Nature Astronomy, 5,
  1017, \dodoi{10.1038/s41550-021-01417-w}

\bibitem[{{Kraft} {et~al.}(2002){Kraft}, {Forman}, {Jones}, {Murray},
  {Hardcastle}, \& {Worrall}}]{kraft2002}
{Kraft}, R.~P., {Forman}, W.~R., {Jones}, C., {et~al.} 2002, \apj, 569, 54,
  \dodoi{10.1086/339062}

\bibitem[{{Krimm} {et~al.}(2013){Krimm}, {Holland}, {Corbet}, {Pearlman},
  {Romano}, {Kennea}, {Bloom}, {Barthelmy}, {Baumgartner}, {Cummings},
  {Gehrels}, {Lien}, {Markwardt}, {Palmer}, {Sakamoto}, {Stamatikos}, \&
  {Ukwatta}}]{krimm2013}
{Krimm}, H.~A., {Holland}, S.~T., {Corbet}, R.~H.~D., {et~al.} 2013, \apjs,
  209, 14, \dodoi{10.1088/0067-0049/209/1/14}

\bibitem[{{Lohfink} {et~al.}(2013){Lohfink}, {Reynolds}, {Jorstad}, {Marscher},
  {Miller}, {Aller}, {Aller}, {Brenneman}, {Fabian}, {Miller}, {Mushotzky},
  {Nowak}, \& {Tombesi}}]{lohfink2013}
{Lohfink}, A.~M., {Reynolds}, C.~S., {Jorstad}, S.~G., {et~al.} 2013, \apj,
  772, 83, \dodoi{10.1088/0004-637X/772/2/83}

\bibitem[{{Mahadevan}(1997)}]{mahadaven1997}
{Mahadevan}, R. 1997, \apj, 477, 585, \dodoi{10.1086/303727}

\bibitem[{{Markowitz} {et~al.}(2003){Markowitz}, {Edelson}, {Vaughan},
  {Uttley}, {George}, {Griffiths}, {Kaspi}, {Lawrence}, {McHardy}, {Nandra},
  {Pounds}, {Reeves}, {Schurch}, \& {Warwick}}]{markowitz2003}
{Markowitz}, A., {Edelson}, R., {Vaughan}, S., {et~al.} 2003, \apj, 593, 96,
  \dodoi{10.1086/375330}

\bibitem[{{Markowitz} {et~al.}(2007){Markowitz}, {Takahashi}, {Watanabe},
  {Nakazawa}, {Fukazawa}, {Kokubun}, {Makishima}, {Awaki}, {Bamba}, {Isobe},
  {Kataoka}, {Madejski}, {Mushotzky}, {Okajima}, {Ptak}, {Reeves}, {Ueda},
  {Yamasaki}, \& {Yaqoob}}]{markowitz2007}
{Markowitz}, A., {Takahashi}, T., {Watanabe}, S., {et~al.} 2007, \apj, 665,
  209, \dodoi{10.1086/519271}

\bibitem[{{Marscher} {et~al.}(2018){Marscher}, {Jorstad}, {Williamson},
  {L{\"a}hteenm{\"a}ki}, {Tornikoski}, {Hunter}, {Leidig}, {Zain Mobeen},
  {Vera}, \& {Chamani}}]{marscher2018}
{Marscher}, A.~P., {Jorstad}, S.~G., {Williamson}, K.~E., {et~al.} 2018, \apj,
  867, 128, \dodoi{10.3847/1538-4357/aae4de}

\bibitem[{{McHardy} {et~al.}(2004){McHardy}, {Papadakis}, {Uttley}, {Page}, \&
  {Mason}}]{mchardy2004}
{McHardy}, I.~M., {Papadakis}, I.~E., {Uttley}, P., {Page}, M.~J., \& {Mason},
  K.~O. 2004, \mnras, 348, 783, \dodoi{10.1111/j.1365-2966.2004.07376.x}

\bibitem[{{M{\"u}ller} {et~al.}(2014){M{\"u}ller}, {Kadler}, {Ojha}, {Perucho},
  {Gro{\ss}berger}, {Ros}, {Wilms}, {Blanchard}, {B{\"o}ck}, {Carpenter},
  {Dutka}, {Edwards}, {Hase}, {Horiuchi}, {Kreikenbohm}, {Lovell}, {Markowitz},
  {Phillips}, {Pl{\"o}tz}, {Pursimo}, {Quick}, {Rothschild}, {Schulz},
  {Steinbring}, {Stevens}, {Tr{\"u}stedt}, \& {Tzioumis}}]{muller2014}
{M{\"u}ller}, C., {Kadler}, M., {Ojha}, R., {et~al.} 2014, \aap, 569, A115,
  \dodoi{10.1051/0004-6361/201423948}
  
\bibitem[{{Mundo}(2022)}]{mundo2022}
{Mundo}, et al.~2022 (in preparation)  

\bibitem[{{Narayan}(1996)}]{narayan1996}
{Narayan}, R. 1996, \apj, 462, 136, \dodoi{10.1086/177136}

\bibitem[{{Neumayer}(2010)}]{neumayer2010}
{Neumayer}, N. 2010, \pasa, 27, 449, \dodoi{10.1071/AS09080}

\bibitem[{{Papadakis}(2004)}]{papadakis2004}
{Papadakis}, I.~E. 2004, \mnras, 348, 207,
  \dodoi{10.1111/j.1365-2966.2004.07351.x}

\bibitem[{{Rani} {et~al.}(2014){Rani}, {Krichbaum}, {Marscher}, {Jorstad},
  {Hodgson}, {Fuhrmann}, \& {Zensus}}]{rani2014}
{Rani}, B., {Krichbaum}, T.~P., {Marscher}, A.~P., {et~al.} 2014, \aap, 571,
  L2, \dodoi{10.1051/0004-6361/201424796}

\bibitem[{{Rothschild} {et~al.}(2011){Rothschild}, {Markowitz}, {Rivers},
  {Suchy}, {Pottschmidt}, {Kadler}, {M{\"u}ller}, \& {Wilms}}]{rothschild2011}
{Rothschild}, R.~E., {Markowitz}, A., {Rivers}, E., {et~al.} 2011, \apj, 733,
  23, \dodoi{10.1088/0004-637X/733/1/23}

\bibitem[{{Rothschild} {et~al.}(2006){Rothschild}, {Wilms}, {Tomsick},
  {Staubert}, {Benlloch}, {Collmar}, {Madejski}, {Deluit}, \&
  {Khandrika}}]{rothschild2006}
{Rothschild}, R.~E., {Wilms}, J., {Tomsick}, J., {et~al.} 2006, \apj, 641, 801,
  \dodoi{10.1086/500534}

\bibitem[{{Shimizu} \& {Mushotzky}(2013)}]{shimizu2013}
{Shimizu}, T.~T., \& {Mushotzky}, R.~F. 2013, \apj, 770, 60,
  \dodoi{10.1088/0004-637X/770/1/60}

\bibitem[{{Vaughan} {et~al.}(2003){Vaughan}, {Edelson}, {Warwick}, \&
  {Uttley}}]{vaughan2003}
{Vaughan}, S., {Edelson}, R., {Warwick}, R.~S., \& {Uttley}, P. 2003, \mnras,
  345, 1271, \dodoi{10.1046/j.1365-2966.2003.07042.x}

\bibitem[{{Vaughan} {et~al.}(2005){Vaughan}, {Fabian}, \&
  {Iwasawa}}]{vaughan2005a}
{Vaughan}, S., {Fabian}, A.~C., \& {Iwasawa}, K. 2005, \apss, 300, 119,
  \dodoi{10.1007/s10509-005-1211-x}

\bibitem[{{Worrall} {et~al.}(2008){Worrall}, {Birkinshaw}, {Kraft}, {Sivakoff},
  {Jord{\'a}n}, {Hardcastle}, {Brassington}, {Croston}, {Evans}, {Forman},
  {Harris}, {Jones}, {Juett}, {Murray}, {Nulsen}, {Raychaudhury}, {Sarazin}, \&
  {Woodley}}]{worral2008}
{Worrall}, D.~M., {Birkinshaw}, M., {Kraft}, R.~P., {et~al.} 2008, \apjl, 673,
  L135, \dodoi{10.1086/528681}

\bibitem[{{Wykes} {et~al.}(2015){Wykes}, {Hardcastle}, \&
  {Croston}}]{wykes2015}
{Wykes}, S., {Hardcastle}, M.~J., \& {Croston}, J.~H. 2015, \mnras, 454, 3277,
  \dodoi{10.1093/mnras/stv2187}

\bibitem[{{Yi} \& {Boughn}(1999)}]{yi1999}
{Yi}, I., \& {Boughn}, S.~P. 1999, \apj, 515, 576, \dodoi{10.1086/307041}

\end{thebibliography}

\end{document}